\documentclass{ws-p10x7}
\def\d0{D_2^{*0}}
\def\d+{D_2^{*+}}
\def\mev{ \,{\rm MeV}/c^2  }
\begin{document}

\title{Results on charmed meson spectroscopy from FOCUS }

 \author{Franco L. Fabbri }
 \address{Laboratori Nazionali di Frascati  via E.Fermi, 40 - Frascati
         (Rome) -  00044 Italy \\
 {\sc   on behalf of the FOCUS(E831) Collaboration}
 }
\twocolumn[\maketitle\abstract{
We report the preliminary measurement by the FOCUS Collaboration
 (E831 at Fermilab) of masses and widths of the 
L=1 charm mesons: a $D_2^{*0}$
state of mass (width) $2463.5 \pm 1.5 \pm 1.5 (30.5 \pm 1.9 \pm 3.8) \mev$
decaying to $D^+\pi^-$, and a $D_2^{*+}$ state of mass (width) $2468.2 \pm
1.5 \pm 1.4 (28.6 \pm 1.3 \pm 3.8) \mev$ 
decaying to $D^0\pi^+$. The fit of the invariant mass distribution
requires an additional  term to account for a broad structure over background.
}]
%
%\section{Excited Charm Mesons}
%
 In this paper we present preliminary results from the FOCUS experiment
\footnote{Coauthors: J.Link, V.S. Paolone, M. Reyes, P.M. Yager
({\bf UC DAVIS}); J.C. Anjos, 
 I. Bediaga, C. G\"obel, J. Magnin, J.M. de Miranda, I.M. Pepe
% ...
} 
(E831 at Fermilab)
on the spectroscopy of bound states of a charm quark and a lighter quark
with orbital angular momentum  $L=1$, called $D_2^* (c\bar u, c\bar d)$.
 With the high-statistics, high-mass resolution experiments attaining
 maturity, emphasys has been shifted from the ground state ($0^-$ and $1^-$)
 $c\bar q$ mesons and $(1/2^+$ and $3/2^+)$ $ cqq$ baryons to the orbitally-
 and, only very recently, radially-excited states\footnote{In the past,
 these excited states were called generically and
 improperly  $D^{**}$.}.
 A consistent theoretical framework for the spectrum of
 heavy-light mesons is given by the ideas of Heavy Quark Symmetry (HQS),
 later generalized by Heavy Quark Effective Theory in the QCD framework.
 The basic idea (mediated from the $JJ$ coupling in atomic
 physics) is that in the limit of infinite heavy quark mass: a) the much
 heavier quark does not contribute to the
 orbital degrees of freedom, which are completely defined by the light
 quark(s) only; and b) properties are independent of heavy quark flavor.
    Heavy Quark Symmetry provides explicit predictions on the
    spectrum of excited charmed
 states\cite{specpap}. 
\par
%
%\begin{figure}%1
%\epsfxsize120pt
%\figurebox{120pt}{160pt}{D-sigsb.eps}
%\caption{Masses and transitions predicted for the excited D meson states.}
%\label{fig:spec}
%\end{figure}
%
    In the limit of infinite heavy
    quark mass, the spin of the heavy quark ${\bf S_Q}$ decouples from the
    light     quark degrees of freedom (spin ${\bf s_q}$ and orbital ${\bf
    L}$), with ${\bf S_Q}$ and ${\bf j_q} \equiv {\bf s_q} +  {\bf
    L}$ the conserved quantum numbers. Predicted
    excited states are formed
    by combining ${\bf S_Q}$ and ${\bf j_q}$. For $L=1$ we have $j_q=1/2$ and
    $j_q=3/2$ which, combined with  $S_Q$, provide prediction for two
    $j_q=1/2$ (J=0,1) states, and two $j_q=3/2$ (J=1,2) states. These
    four states
    are named respectively $D_0^*$, $D_1(j_q=1/2)$,
    $D_1(j_q=3/2)$ and $D_2^*$.
   Finally, parity and
    angular momentum
    conservation favor the $(j_q=1/2)$ states to decay to the ground
    states mainly via S-wave transitions (broad width), while  $(j_q=3/2)$
    states 
    would     decay via D-wave (narrow width). While the narrow states are
 well established, the evidence for the broad states (both in the $c$-quark
 and in the $b$-quark sector) is much less stringent \cite{reviews}.
%
% FOCUS
%
\par
The data for this paper were collected in the Wideband photoproduction
experiment FOCUS during the Fermilab 1996--1997 fixed-target
run. FOCUS is a considerably upgraded version of a previous experiment, 
E687 \cite{Frabetti:1992au}. In FOCUS, a forward multi-particle
spectrometer is used to  
measure the interactions of high energy photons on a segmented BeO
target. We obtained a sample of over 1 million fully reconstructed
charm particles in the three major decay modes: $D^0 \rightarrow K^- \pi^+
,~K^- \pi^+ \pi^- \pi^+$ and $D^+ \rightarrow K^- \pi^+ \pi^+$.  
\par
The FOCUS detector is a large aperture, fixed-target spectrometer with
excellent vertexing, particle identification, and reconstruction
capabilities for photons and $\pi^0$'s. A photon beam is
derived from the bremsstrahlung of secondary electrons and positrons
with an $\approx 300$ GeV endpoint energy produced from the 800
GeV/$c$ Tevatron proton beam. The charged particles which emerge from
the target are tracked by two systems of silicon microvertex
detectors. The upstream system, consisting of 4 planes (two views in 2
stations), is interleaved with the experimental target, while the
other system lies downstream of the target and consists of twelve
planes of microstrips arranged in three views. These detectors provide
high resolution separation of primary (production) and secondary
(decay) vertices with an average proper time resolution of $\approx
30~ {\rm fs}$ for 2-track vertices. The momentum of a charged particle
is determined by measuring its deflections in two analysis magnets of
opposite polarity with five stations of multiwire proportional
chambers. Three multicell threshold \v Cerenkov counters are used to
discriminate between pions, kaons, and protons.  
%
%\section{Analysis Procedure and Results}%1
%
\par
\begin{figure}[ht]
 \vspace{10cm}
 \includegraphics{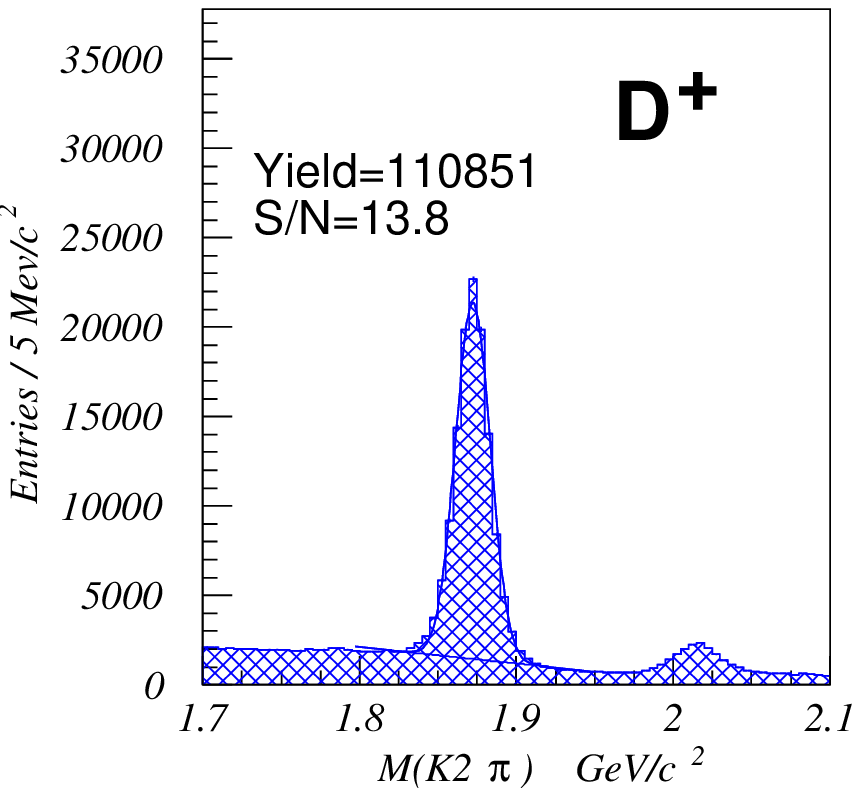}
 \includegraphics{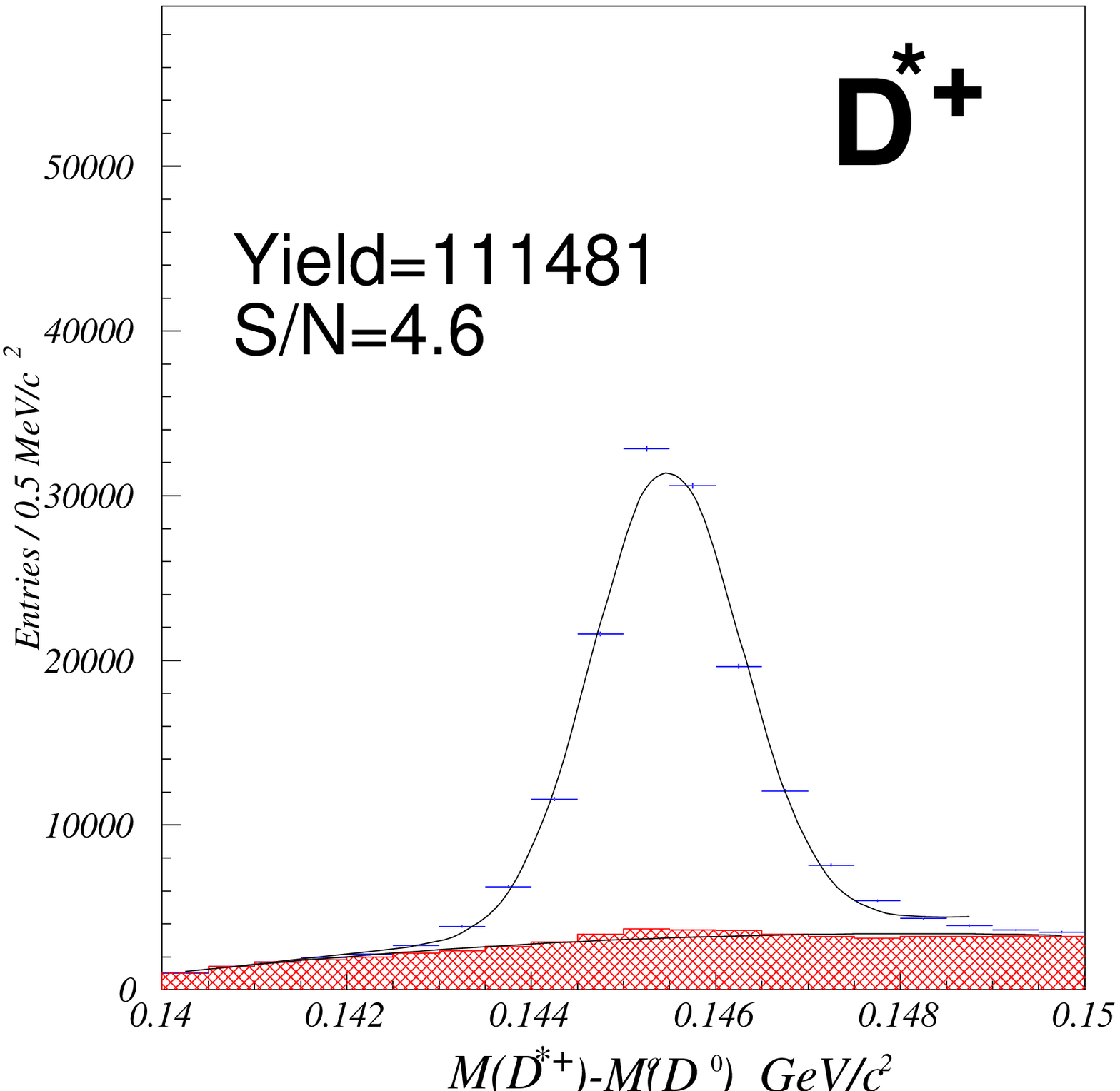}
 \caption{Mass plots for $D^+\rightarrow K^-\pi^+ \pi^+$ (top); and 
 $D^{*+} \rightarrow D^0\pi^+$ (bottom).
    \label{fig:d0} }
\end{figure}

The decays 
 $D^0\rightarrow K^- \pi^+ $, 
 $D^+\rightarrow K^- \pi^+\pi^+$, 
 $D^0\rightarrow K^- \pi^+ \pi^- \pi^+$, 
 $D^{*+} \rightarrow D^0 \pi^+$
were selected (Fig.\ref{fig:d0}). To ensure clean charm samples, candidate
events were selected with a large $\ell/\sigma_\ell$, being  $\ell$ the
separation of the primary and weak 
decay vertices, and $\sigma_\ell$ its 
uncertainty; the primary multiplicity was required to be greater than 1,
and the primary to be located within one of the interaction targets; the
kaon and pion candidates to be  consistent with the kaon and pion
hypotesis, based on the \^Cerenkov identification system;
the weak decay vertex to be outside of the interaction targets 
$(\sigma_{OoM}>0)$ for the $D^0\rightarrow K^- \pi^+ \pi^- \pi^+$ mode
only; and $| \cos\theta_K| < 0.7$ for the $D^0\rightarrow K^- \pi^+ $ mode
only, where the decay angle $\theta_K$ is defined as the angle which, in
the $D$ rest frame,  the kaon momentum forms with the $D$ momentum in the
lab. The results in this paper have been shown to be insensitive to the
detailed choice of selection parameters. 
\par
\begin{figure*}
%\epsfxsize30pc
%\figurebox{21pc}{32pc}{}
 \vspace{9.1cm}
  \includegraphics{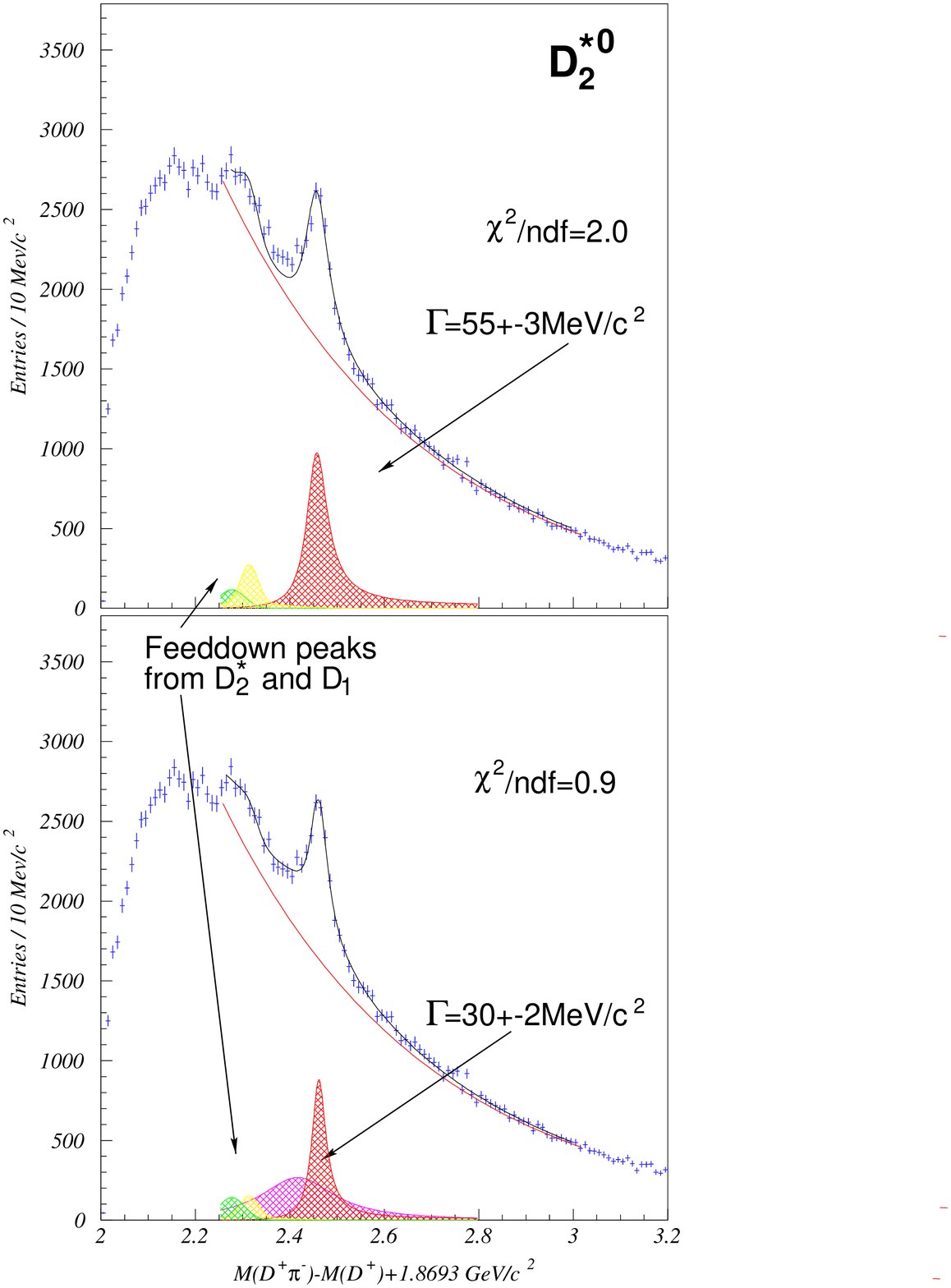}
  \includegraphics{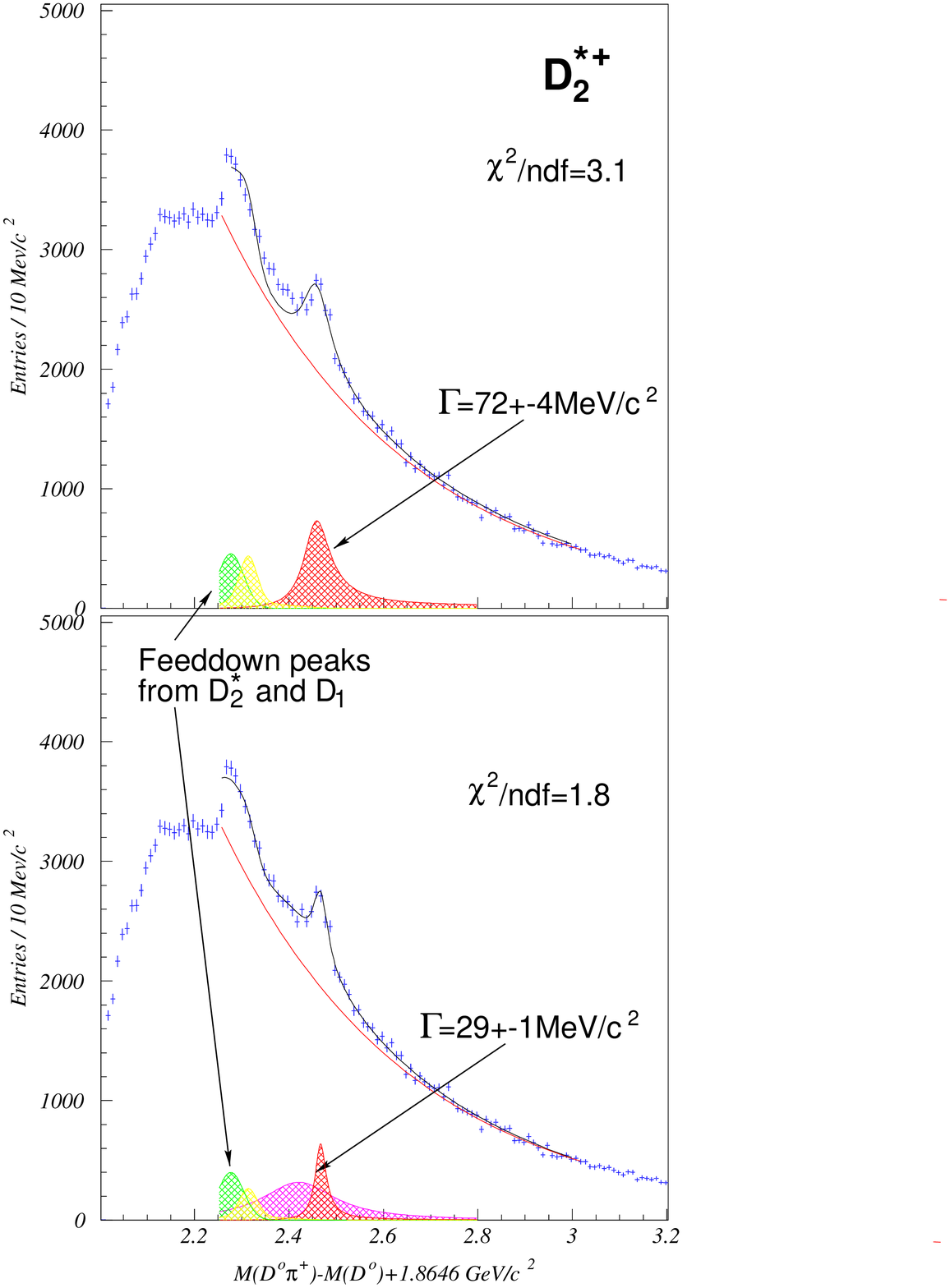}
\caption{
 The $D^+\pi^-$ ($D^0\pi^+$) mass spectra is shown on the left (right).
 The invariant mass variable is defined as
 $\Delta M_0\equiv M(D^+\pi^-) - M(D^+) + M_{PDG}(D^+)$ and 
  $\Delta M_+\equiv M(D^0\pi^+) - M(D^0) + M_{PDG}(D^0)$, respectively.
%
% In each of the top plots, the mass spectrum was fitted 
% with a relativistic $D$-wave Breit-Wigner function, convoluted with a
% gaussian resolution function, and the sum of an exponential and two
% gaussians for the background.
% For each final state, the bottom histogram shows the fit results when
% a $S$-wave relativistic Breit-Wigner function is added to the above 
% fit functions.
%
\label{fig:d0+-}}
\end{figure*}

The principal preliminary result in this
 paper relates to a study of the $D^+\pi^-$ and   $D^0\pi^+$ mass spectra.
The $D^+$ or $D^0$ candidates were combined with the pion tracks in the primary
vertex to form L=1 $D$-meson candidates. Events with charm candidates
 coming from $D^*$ decays were rejected by applying 
 a $\pm 3 \sigma$ cut around the $D^*-D$ mass difference.
Figure~\ref{fig:d0+-} shows the distribution in the invariant mass
 difference 
 $$\Delta M_0\equiv M(D^+\pi^-) - M(D^+) + M_{PDG}(D^+).$$
 The plot shows a pronounced peak, consistent with being due to
 a $D_2^{*0}$ of mass  $M \approx 2460 \mev$. Because of the narrow width,
 this state has traditionally been identified as the $J=2^+$ state. The
 additional enhancement at $M\approx 2300 \mev$ is consistent, as verified
 from Monte Carlo simulations, with arising from the feed-down of the states
 $D_1^0$ and $D_2^{*0}$ decaying to $D^{*+}\pi^-$, with 
 the $D^{*+}$ subsequently decaying to $D^+$ and undetected neutral pion.
% \par
 The $D_2^{*0}$ signal was fitted with a relativistic $D$-wave
 Breit-Wigner function, convoluted with a gaussian resolution function
 $(\sigma = 7 {\rm MeV})$. The background was fitted with the  sum of an
 exponential, and two gaussians for the feed-downs described above, whose
 peaks and widths were fixed at the Monte Carlo values. The
 slope of the exponential was fixed to the value determined by a fit to the
 wrong-side events  mass distribution, which is very well described by a
 single-slope  exponential  in the entire fitting interval $2250-3000 \mev$.
 For this fit we get a $\chi^2/{\tt dof} =2$, and a $\Gamma = 55\pm 3 \mev$
 $D_2^{*0}$ width statistically non compatible with the PDG2000 world
 average of $\Gamma = 23\pm 5 \mev$. We then add an  $S$-wave relativistic
 Breit-Wigner function  to the fit, which improves the fit quality
 $\chi^2/{\tt dof} =0.9$, and provides a width $\Gamma = 30\pm 2 \mev$
 compatible to the PDG2000 value.
 \par
 The mass difference 
$$\Delta M_+\equiv M(D^0\pi^+) - M(D^0) + M_{PDG}(D^0)$$
 spectrum (Fig.~\ref{fig:d0+-} shows structures similar to those in the
 $\Delta M_0$ spectrum. The prominent peak is consistent with being due to
 a $D_2^{*+}$ of mass  $M \approx 2460 \mev$. 
 The additional enhancement at $M\approx 2300 \mev$ is consistent, as verified
 from Monte Carlo simulations, with arising from the feed-down of the states
$D_1^0$ and $D_2^{*0}$ decaying to $D^{*0}\pi^+$, with 
 the $D^{*0}$ subsequently decaying to $D^0$ and undetected neutral pion.
 The fitting procedure for the $\Delta M_+$ spectrum follows the same
 guidelines as the $\Delta M_0$. Several systematics checks have been
 performed to verify the stability of our measurements of masses and
 widths, such as fit variants varying the selection cuts over an extended
 range, and the stability of our mass measurements when performed on
 statistically independent subsamples ($\ell /\sigma_\ell$ greater and less
 than 30, particle vs. antiparticle, momentum of the pion from the
 $D_2^*$ decay greater and less than $18 {\rm GeV} /c$, momentum of the $D$
 meson greater and less than  $70 {\rm GeV} /c$).
 Table~\ref{tab:mass} summarizes the preliminary results on the 
measurements of masses and widths from
 the study of $D^0\pi^+$ and $D^+\pi^-$ final states.
%\begin{figure}[t]
% \epsfig{file=prova.eps,width=12cm,height=4cm} 
% \epsfig{file=osakap13.eps,width=8cm,height=10cm} 
% \caption{Observation of \label{fig:d0+-} }
%\end{figure}
%
%
\par
 \begin{table}
 \caption{Preliminary measurements of masses and widths for narrow
 structures in   $D^+\pi^-$ and $D^0\pi^+$ invariant mass spectra. 
   \label{tab:mass}
 }
 \footnotesize
 \begin{center}
\begin{tabular}{|l|c|c|} \hline
             &  Mass                     & Width                 \\
             & $\mev$                    & $\mev$                \\
\hline
 $D_2^{*0}$  &  $2463.5 \pm 1.5 \pm 1.5$ & $30.5 \pm 1.9 \pm 3.8$ \\
  PDG2000    &  $2458.9 \pm 2.0$         & $23 \pm 5$             \\
\hline
 $D_2^{*+}$  &  $2468.2 \pm 1.5 \pm 1.4$ & $28.6 \pm 1.3 \pm 3.8$ \\
  PDG2000    &  $2459\pm4$               & $25^{+8}_{-7}$         \\
\hline
\end{tabular}
  \vfill
 \end{center}
\end{table}

\par
 In conclusion, FOCUS has collected  a large sample of $D^{*0}_2$ and 
$D^{*+}_2$ L=1 mesons out of the total sample of about $10^6$ D meson
states, and $0.5\,10^4$ $D^*$ meson states. The study of the $D\pi$ mass
spectrum provides new preliminary values of the masses and widths for the
$D_2^*$ meson (Tab.~\ref{tab:mass}). The $D\pi$ mass spectrum (once
subtracted the background, the $D_2^*$ signal, and the expected feed-downs)
shows an excess of events centered around $2420 \mev$ and about $185 \mev$
wide. A broad $(\sim 100-200 \mev)$ state (the $D_0^*$) is predicted by HQS
at about $2350 \mev$. The observed excess could be reminiscent of this
state, or of a feed-down from another broad state such as the $D_1
(j_q=1/2)$, possibly interfering.
Work is in progress to verify such hypothesis.  
%
%\section*{Acknowledgments}
%
\par
We wish to acknowledge the assistance of the staffs of Fermi National
Accelerator Laboratory, the INFN of Italy, and the physics departments
of the collaborating institutions. This research was supported in part
by the U.~S.  National Science Foundation, the U.~S. Department of
Energy, the Italian Istituto Nazionale di Fisica Nucleare and
Ministero dell'Universit\`a e della Ricerca Scientifica e Tecnologica,
the Brazilian Conselho Nacional de Desenvolvimento Cient\'{\i}fico e
Tecnol\'ogico, CONACyT-M\'exico, the Korean Ministry of Education, and
the Korean Science and Engineering Foundation.

\end{document}